\documentstyle[11pt,aaspp4,epsf]{article}
\newcommand{\vect}[1]{\mbox{\boldmath${#1}$}}

\newcommand{\lmk}{\left(}
\newcommand{\rmk}{\right)}
\newcommand{\lnk}{\left\{ }
\newcommand{\rnk}{\right\} }
\newcommand{\lkk}{\left[}
\newcommand{\rkk}{\right]}
\newcommand{\lla}{\left\langle}
\newcommand{\rra}{\right\rangle}

\newcommand{\calv}{{{\cal V}_L}}
\newcommand{\vex}{{\vect x}}
\newcommand{\vey}{{\vect y}}
\newcommand{\veV}{{\vect V}}

\newcommand{\beq}{\begin{equation}}
\newcommand{\eeq}{\end{equation}}
\newcommand{\beqa}{\begin{eqnarray}}
\newcommand{\eeqa}{\end{eqnarray}}
\newcommand{\hMpc}{h^{-1}{\rm Mpc}}

\newcommand{\hV}{\hat{V}_R}
\newcommand{\Pp}{\Phi_{\scriptsize\parallel}}
\newcommand{\Pv}{\Phi_{\scriptsize\perp}}
\newcommand{\etal}{et al.\ }
\begin{document}

\begin{minipage}[c]{10cm}
\vspace{0.5cm}
YITP-98-9,  SU-ITP 98/03, KUNS-1491\\
\end{minipage}\\

\title{Sample variance of the cosmic velocity field \\}
\author{\sc Naoki Seto}
\affil{Department of Physics, Faculty of Science, Kyoto University,
Kyoto 606-01, Japan
%\\seto@tap.scphys.kyoto-u.ac.jp
}

\author{\sc Jun'ichi Yokoyama}
\affil{Department of Physics, Stanford University, Stanford
CA94305-4060\\
 and
Yukawa Institute for Theoretical Physics, Kyoto University,
Kyoto 606-01, Japan
%\\yokoyama@yukawa.kyoto-u.ac.jp
}

%\date{\today}
\abstract{Since the cosmic peculiar velocity field depends on
small wave-number modes strongly, we cannot probe its universal
properties unless we observe a sufficiently large region.
We calculate the expected deviation (sample variance) of the
peculiar velocity dispersion from its universal value in the case
observed volume is finite.
Using linear theory we show that the sample variance remains as
large as $\sim 10\%$, even if the observed region is
as deep as $100\hMpc$ and that it seriously affects the estimation
of cosmological parameters from the peculiar velocity field.
\\
{\it Subject headings:}
cosmology: theory --- large-scale structure of the Universe
--- methods: statistical}

\section{Introduction}
Observational analysis of the large-scale peculiar velocity field is
considered as a very
effective method to impose  constraint on the cosmological parameters as well
as the spectrum of the primordial density fluctuations in the universe.
Since its power spectrum depends more strongly  on small wave-number
modes, which are less contaminated by nonlinear effects,
 than that of density fluctuations, linear perturbation theory suffices to
reproduce the amplitude of the peculiar velocity even on a relatively
small length
scales.  In fact, Bahcall et al. (1994) have calculated the one-point
peculiar-velocity dispersion  smoothed with a Gaussian filter over the
smoothing
scale $3\hMpc$  in their large-scale $N-$body simulations with various
cosmological models and have found an excellent agreement between their
calculations and the predictions of linear theory.  Here $h$ is as usual
the present Hubble parameter in unit of 100km/sec/Mpc.
Validity of the linear analysis has also been confirmed
 analytically in the framework of higher order Eulerian
  perturbation theory by Makino, Sasaki \& Suto (1992),
who showed that
second-order effects were negligible on the smoothed
velocity dispersion in the case of cold-dark-matter power spectrum.

The stronger dependence on the smaller wave-number modes, on the other hand,
implies that the correlation length of peculiar velocity field is larger
than that
of the density field and that
we must survey a volume large enough to contain sufficient number
of such modes.  In that sense, Bahcall et al. (1994) was quite correct
in adopting
a simulation box whose dimension was as large as $800\hMpc$.
At present, however, we can by no means hope to observe the peculiar velocity
field that far in the real Universe.
For example, the observed depth of the
recent high quality catalogue by Giovanelli et al. (1997) has been limited to
$100\hMpc$ mainly due to the difficulties to estimate the distance which is
of course essential in obtaining the line-of-sight peculiar
velocity (see also Giovanelli et al. 1996).
Thus we should be careful in interpreting the observational data to
compare with
theoretical predictions of various cosmological models.

In this Letter we present a simple analysis of the uncertainty caused by the
finiteness of the sampled volume, which we call the sample variance.  We
concentrate on the one-dimensional velocity dispersion of the smoothed
peculiar velocity field which is the sole quantity to characterize its
distribution
on scales linear theory suffices, provided that primordial fluctuations are
Gaussian distributed, which we assume throughout the paper.
Uncertainties  related to the determination of
the peculiar velocity dispersion include
various factors, such as the observational errors in the estimation of
distances to the objects,  sparseness of the number
of the clusters/galaxies
probed, and so on.
Monte Carlo calculations using mock catalogues including all these errors
have been done in the literatures ({\it e.g.} Borgani \etal 1997),
which are
very complicated and require high computational costs.
In contrast to these approaches,
our purpose in this article is to clarify the fundamental limitations due
to the finiteness of the sample, which is very important in the sense that
the uncertainty is independent of  how accurate we could measure the
peculiar velocities of galaxies or clusters.

We develop a  simple analytical formula similar to the same kind
of analysis on the CMB measurement (Scott \etal 1993).
We show that the sample variance causes an uncertainty
 as large as $10\%$ on the
measurement of velocity dispersion, even if we take a full
sphere of radius  $100\hMpc$ around us as the observational volume.
This is comparable to the current observational error (Bahcall \&Oh 1996).

\section{Formulation}
%\vspace{1.0cm}

First we define the peculiar velocity field, $\veV (\vex)$, smoothed
over a radius
$R$ by
\beq
\veV_R(\vex)=\int \veV(\vex)W_R(\vex-\vex')d^3x',
\eeq
where $W_R(\vex-\vex')$ is the Gaussian window function given by
$W_R(\vex)=(2 \pi)^{-3/2}R^{-3} \exp\lmk -  {\vex^2}/{2R^2}\rmk$.
We define the  line-of-sight component by
\beq
\hV(\vex)=\veV_R(\vex)\cdot \frac{\vex}{|\vex|}.
\eeq
It is true that, in reality, we must first project and then smooth, but we
ignore this incommutability since we consider a survey region
whose dimension is much larger than $R$.  Otherwise the statistical average
would become nonsense.  To be specific we assume that the observed region is
a sphere with radius $L$ and denote its volume by
$\calv\equiv \frac{4}{3}\pi L^3$ with $L \gg R$.

One dimensional velocity dispersion estimated from the line-of-sight
component in the surveyed volume $\calv$ is written as
\beq
X(L,R) \equiv \frac1{\calv}\int_\calv \hV(\vex)^2d^3x.
\eeq
The ensemble average of this dispersion is of course independent of the size
of the observed volume, and it is simply the one-dimensional velocity
dispersion with
smoothing radius $R$, which is denoted as
\beq
 \lla X(L,R) \rra =\sigma^2_{1D}(R).
\eeq
Here and hereafter, the angular bracket represents an average
over ensembles of the "universes".
The error associated with the estimation of $X(L,R)$ due to the finiteness of
the sampled volume is characterized by the variance,
\beq
\lla \lkk X(L,R)-\lla X(L,R)\rra\rkk^2\rra=\lla X^2(L,R)\rra - \lla
X(L,R)\rra^2.
 \eeq

We assume that primordial density fluctuation is isotropic random Gaussian,
and apply linear theory, since we adopt  large enough smoothing
scales suggested by Bahcall \etal (1994).
Using properties of the Gaussian distribution, we obtain
\beqa
\lla X^2(L,R) \rra &=& \frac1{\calv^2} \int_\calv d^3x\int_\calv d^3y
\lla \hV(\vex)^2
\hV(\vey)^2 \rra     \nonumber \\
&=&\frac1{\calv^2} \int_\calv d^3x\int_\calv d^3y \lmk \lla \hV(\vex)^2 \rra
\lla\hV(\vey)^2 \rra+2\lla\hV(\vex)\hV(\vey) \rra^2 \rmk. \nonumber
\eeqa
Hence the variance of $X(L,R)$ reads
\beq
\lla X^2(L,R) \rra-\lla X(L,R) \rra^2= \frac2{\calv^2} \int_\calv
d^3x\int_\calv d^3y
\lla\hV(\vex)\hV(\vey) \rra^2. \label{g1}
\eeq
Here
$\lla\hV(\vex)\hV(\vey) \rra$ depends  only on the geometry
 decided by $\vex$ and $ \vey$, namely, on the three variables
$|\vex|=r_1, |\vey|=r_2$ and
the angle $\theta$ between $\vex$ and $\vey$.
Let us normalize the above dispersion by $\lla X(L,R) \rra^2=\sigma^4_{1D}(R)
=\lla \hV(\vex)\hV(\vex) \rra^2$
to define
\beq
E(L,R)\equiv \frac{\lla X^2(L,R) \rra-\lla X(L,R) \rra^2}{\lla X(L,R)
  \rra^2}=\frac{16\pi^2}{\calv^2}\int_0^Ldr_1
\int_0^Ldr_2 \int_0^{\pi}d\theta r_1^2 r_2^2 \sin\theta f(r_1,r_2,\theta)^2,
\eeq
with
\beq
f(r_1,r_2,\theta)\equiv\frac{\lla\hV(\vex)\hV(\vey)
  \rra}{\lla\hV(\vex)\hV(\vex) \rra}. \label{g2}
\eeq

Next we calculate the correlation function
  $f(r_1,r_2,\theta)$. After some algebra, we can relate it with parallel and
  perpendicular
velocity covariance functions,  $\Pp$ and $\Pv$, defined by G\'orski (1988) as
\beq
f(r_1,r_2,\theta)=\Pp(r_{12})(n_x^2\cos\theta+n_xn_y\sin\theta)
 +\Pv(r_{12})\lnk \cos\theta n_y^2-\sin\theta n_xn_y\rnk,
\eeq
where, $r_{12}$, $n_x$ and $, n_y$ are
defined by
\beqa
r_{12}^2=|\vex-\vey|^2=r_1^2+r_2^2-2r_1r_2\cos\theta,~~
n_x=\frac{r_1 \cos\theta-r_2}{r_{12}},~~
n_y=\frac{r_1 \sin\theta}{r_{12}}.
\eeqa
$\Pp (r)$ and $\Pv(r)$ are the velocity correlation functions normalized to
unity at $r=0$, or
 $\Pp(0)=\Pv(0)=1$, and they are written in terms of the  power
spectrum of linear density fluctuations, $P(k)$,
as follows (G\'orski 1988).
\beqa
\Pp(r)&\propto&
\int_0^{\infty} dkP(k) W_R(k)^2  \lmk j_0(kr)-2\frac{j_1(kr)}{kr} \rmk, \\
\Pv(r) &\propto& \int_0^{\infty} dk P(k) W_R(k)^2  \frac{j_1(kr)}{kr}.
\eeqa
In the above expressions  $W_R(k)\equiv \exp(-k^2R^2/2)$
 is the Fourier transform of $W_R(\vex)$,
and $j_m(z)$ is the spherical Bessel functions of the $m-$th order.

In the same manner we can also calculate the sample variance of density
fluctuations smoothed over a radius $R$ at the level of the linear theory.
In this case we only have to replace the line-of-sight peculiar velocity
$\hV(\vex)$ by the density contrast smoothed over the same radius
$R$, $\delta_R(\vex)$.
Thus $\lla\hV(\vex)\hV(\vey) \rra$ in equation (\ref{g1}) is replaced by
$\lla \delta_R(\vex)\delta_R(\vey)\rra$, and $f$ in  equation (\ref{g2}) is
replaced by $\Xi(r_{12})\equiv {\xi_R(r_{12})}/{\xi_R(0)}$,
where $\xi_R(r)$ is the two point correlation function of the smoothed
density field.

In this article, we adopt the power spectrum of cold-dark-matter (CDM) models
given in Efstathiou, Bond \& White (1992)
as
\beq
 P(k)={Bk}{\lnk 1+\lkk \alpha k+(\beta k)^{3/2}+(\gamma k)^2
   \rkk^{\mu}\rnk^{-2/\mu}}  \label{CDM}
\eeq
where
 $\alpha=(6.4/\Gamma)\hMpc$, $\beta=(3.0/\Gamma)\hMpc$,$\gamma=(1.7/\Gamma
)\hMpc$, $\mu=1.13$, and $B$ is the normalization factor.
Since we are dealing with normalized quantities  in the linear theory,
the system is characterized by the shape
parameter $\Gamma=\Omega_0 h$, and the smoothing radius $R$ only.

\section{Results}

First we plot $\Pp(r), \Pv(r)$, and $ \Xi(r)$ in Figure 1.
Apparently, the peculiar velocity field has  larger correlation length than
the density field, in particular, $\Pv(r)$ has a very large correlation length.
This is understandable because in  linear theory the density
contrast,
$\delta$,
and the peculiar velocity, $\veV$, satisfy the relation
$\veV \propto \nabla \cdot \nabla^{-2}\delta $,
which implies that
 the power spectrum of the peculiar velocity field, $P_v(k)$,
is related to $P(k)$ as
$P_v(k)\propto k^{-2}P(k)$ (Peebles 1980).
Therefore, velocity field  is more weighted to smaller $k$ and hence has a
 much larger correlation length, which causes a larger sample variance as we
see below.

In Figure 2, we plot the sample variance of the
one-dimensional velocity dispersion,
$E^{1/2}(L,R)$, with smoothing radius
$R=3(0.2/\Gamma)\hMpc$ and $6(0.2/\Gamma)\hMpc$.
For comparison we also plot the corresponding sample variance for the
density field in linear theory.
To simplify our presentation, the length scale in the horizontal axis is shown
in unit of $(0.2/\Gamma)\hMpc$.

Apparently, the  error due to the finiteness of the observed volume
is much larger for  the velocity field than for the linear density field
\footnote{Note that linear theory is insufficient to reproduce the magnitude of
density fluctuations on the scales we are dealing with.  Hence we do not claim
the above result reflects the correct sample variance of the density field.
All we would like to stress here is that magnitude of the sample variance
is sensitive to the functional shape of the power spectrum.}.
This figure shows clearly that even with a spherical observed region
with radius
 $100\hMpc$, the
error on the estimation of the one-dimensional velocity dispersion,
which is approximately equal to $E^{1/2}(L,R)/2$,
is as large as 10\%.
This is comparable to the observational error of $\sim10\%$ in Bahcall and
Oh (1996).  Note that magnitude of the error we have obtained here
is for an ideal case that a specific region was observed
fully and homogeneously, and that the actual sample variance would be
even larger with the same depth of the sample.

\section{Effects of the peculiar velocity of the observer}
So far, we have not specified the peculiar velocity of the observer, who is
supposedly at the center of the surveyed sphere with radius $L$.
The peculiar velocity of the observer can be estimated
from the dipole anisotropy of the cosmic microwave background radiation.
If we assume
that it is totally due to our peculiar motion then COBE observation
gives $V_{obs}=627{\rm km/s}$  (Kogut \etal 1993).
One may suspect that if the correlation of the peculiar velocity field is so
strong, the observed velocity dispersion in the sphere may  strongly be
affected by the value of $V_{obs}$.
Here we estimate how much the observed velocity dispersion is expected to
shift as a function of $\nu \equiv V_{obs}/\sigma_{1D}(R)$.

The conditional probability distribution function of the radial peculiar
velocity has
been calculated in G\'orski (1988, eq.\ [7]) with a specific value of
the peculiar velocity
of the observer, from which we find
\beq
  \lla \hV^2(\vect{r}) \rra_\nu = \sigma_{1D}^2(R)
  \lkk 1-\Pp(r)^2(1-\nu^2\cos^2 \Theta)\rkk,
\eeq
where $\lla \cdots \rra_{\nu}$ denotes the ensemble average on
condition that the observer has a peculiar velocity of $|{\vect
V}_{obs}|=\nu\sigma_{1D}(R)$, and $\Theta$ is the angle between
$\veV_{obs}$ and $\vect r$.
We thus find  the constrained ensemble average of $X(L,R)$ is given by
\beq
\lla X(L,R) \rra_{\nu}
=\frac{\sigma_{1D}(R)^2}{\calv}\int_{\calv}d^3r
\{1-\Pp(r)^2(1-\nu^2\cos^2\Theta)\}.
\eeq
Therefore the fractional change of the average due to the additional
constraint is given by
\beq
\frac{\lla X(L,R)\rra_\nu - \lla X(L,R)\rra}{\lla X(L,R)\rra}=
\frac{\nu^2-3}{L^3}\int_0^L dr \Pp(r)^2r^2.
\eeq
For COBE-normalized CDM model (with $h=0.75,~\Gamma=0.2$)
we find $\nu=1.58$ and  the above change is only
about $0.5\%$ for $R=3(0.2/\Gamma)\hMpc$ and $L=100(0.2/\Gamma)\hMpc$,
much smaller than the sample variance we have discussed.
Hence the peculiar motion of the observer does not cause a serious problem.

\section{Discussion}

We have shown that the fact that peculiar velocity field is very sensitive to
small wave-number modes implies that we must observe a sufficiently large
region in order to extract useful information on cosmological models from
observational data.   We have dealt with a smoothed peculiar velocity field,
taking the smoothing length large enough to warrant the validity of  linear
theory in which the only important quantity is the dispersion as long as
the primordial fluctuations are distributed Gaussian.  Hence we concentrated
on the sample variance on the dispersion of
the one-dimensional line-of-sight peculiar velocity
and have shown that it is not negligible at all.

Although estimation of cosmological parameters from the analysis of
peculiar velocity field mostly uses  more contrived methods ({\it e.g.}
Dekel 1994), it would be natural to expect that similar limitation
applies in these approaches as well, because they essentially start with the
same kind of observational measures, namely, the line-of-sight peculiar
velocity, and because
linear theory has been shown to suffice on the relevant scales (Bahcall
\etal 1994).
In order to give a rough idea of magnitude of the error on the
estimation of
cosmological parameters caused by the sample variance, let us take the
smoothing
radius large enough that the linear theory applies for the density
contrast as well,
say $R=12\hMpc$, and  relate the root-mean-square density
 fluctuation obtained from galaxy distribution,
$\sigma_{gal}(R)$,
with the one-dimensional peculiar velocity dispersion $\sigma_{1D}(R)$.
 Using the formula of Colberg
\etal (1997), we find
$   \sigma_{1D}^2(R)=\Omega_0^{1.2}
     \sigma_{gal}(R)^2
   F(\Gamma,R)/b^2 $,
where $b$ is the bias parameter and $F(\Gamma,R)$ is a function decided
only by the shape of the matter power spectrum and $R$.
Even in the ideal and hypothetical  case that both $\sigma_{gal}(R)$ and the
coefficient $F(\Gamma, R)$ are known exactly, the above
simple
formula tells us that we have an inevitable uncertainty in
$\Omega^{0.6}/b$ of about
$25\%$ for $L=60\hMpc$ and $20\%$ for $L=100\hMpc$ (in the case CDM
spectrum (\ref{CDM}) with $\Gamma=0.2$ is adopted).

In conclusion, the strong dependence of the
peculiar velocity field
on small wave-number modes  has the two aspects; the advantage is that the
linear theory suffices even on relatively small scales, while  the
disadvantage
 is
that we must have a larger observational volume to measure it accurately.

\acknowledgments

We are grateful  to Prof.\
N. \ Sugiyama for useful comments. NS would like
to thank  Prof.\ H.\ Sato
for his continuous encouragements.   JY is grateful to Prof.\
Andrei Linde for his hospitality at Stanford University where part of
the work was done and to the Monbusho for financial support.  NS
acknowledges support from JSPS.
This work was partially supported by the Japanese Grant
in Aid for Science Research Fund of the Monbusho
  No.\ 3161(NS) and No.\ 09740334(JY).

\newpage
%\centerline{\bf FIGURES}

\begin{figure}
 \begin{center}
 \epsfxsize=8.2cm
 \begin{minipage}{\epsfxsize} \epsffile{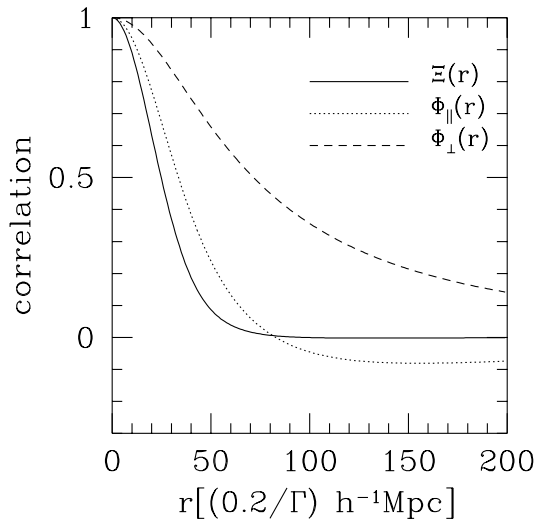} \end{minipage}
 \end{center}
\caption[]{\label{fig:2D}
The normalized correlation functions of velocity and density with
smoothing radius $3(0.2/\Gamma)\hMpc$.
The solid curve represents the two point correlation function of the
linear density field, 
$\Xi(r)$.
The dotted line represents the  parallel velocity covariance
function, $\Pp(r)$, and dashed line  represents the  perpendicular
component, $\Pv(r)$.
}
\end{figure}

\begin{figure}
 \begin{center}
 \epsfxsize=8.2cm
 \begin{minipage}{\epsfxsize} \epsffile{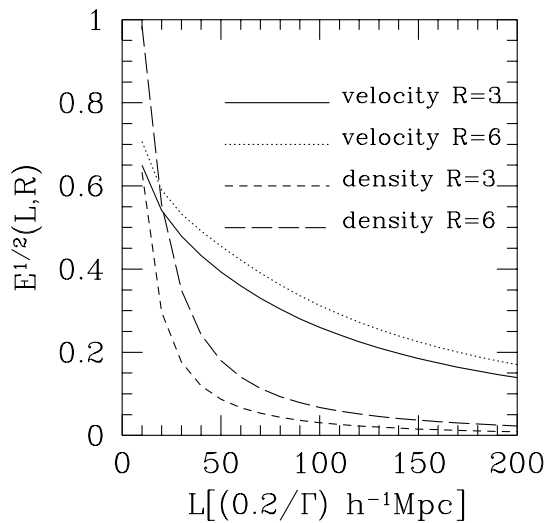} \end{minipage}
 \end{center}
\caption[]{$E^{1/2}(L, R)$, the  expected error on the estimation of
 the universal velocity dispersion.
The horizontal axis is the radius $L$ of the observed sphere
in unit of $(0.2/\Gamma) \hMpc$.
The dotted line corresponds to $E^{1/2}(L,6)$,
and the solid line to $E^{1/2}(L,3)$, both for the
one dimensional peculiar  velocity.

The long-dashed and short-dashed lines represent the corresponding
quantities for the linear density contrast.}
\end{figure}

\end{document}